\documentclass[10pt]{article}
\usepackage{amsmath,amssymb,amsthm,amsfonts}
\usepackage{graphicx}
\usepackage[utf8]{inputenc} % Any characters can be typed directly from the keyboard, eg éçñ
\usepackage{textcomp} % provide lots of new symbols
\usepackage{flafter}  % Don't place floats before their definition
\usepackage{natbib} % use author/date bibliographic citations
\usepackage{bm}  % Define \bm{} to use bold math fonts
\usepackage{caption}

\date{}
\begin{document}

\title{A Bayesian approach to modeling mortgage default and prepayment}
\author{Arnab Bhattacharya, Simon P. Wilson, Refik Soyer} \maketitle

\begin{abstract}
\noindent In this paper we present a Bayesian competing risk proportional hazards model to describe mortgage defaults and prepayments. We develop Bayesian inference for the model using Markov chain Monte Carlo methods. Implementation of the model is illustrated using actual default/prepayment data and additional insights that can be obtained from the Bayesian analysis are discussed.  
\end{abstract}

% \begin{keywords}
% \noindent {\bf Keywords:} % \end{keywords}

\section{Introduction and Overview}

From a legal point of view mortgage default is defined as ``transfer of the legal ownership of the property from the borrower to the lender either through the execution of foreclosure proceedings or the acceptance of a deed in lieu of foreclosure"; see \citet{Gilberto:1989aa}. However, as noted by \cite{Ambrose:1998aa}, it is common in the literature to define default as being delinquent in mortgage payment for ninety days.

There exists a rich literature on modeling mortgage default risk; see for example, \cite{Quercia:1992aa} and \cite{Leece:2004aa}. An important class of models is based on the ruthless default assumption which states that a rational borrower would maximize his/her wealth by defaulting on the mortgage if the market value of the mortgage exceeds the house value, and by prepaying via refinancing if the market value of the house exceeds the book value of the house. Such models use an option theoretic approach and assume that the mortgage value and the prepayment and default options are determined by the stochastic behavior of  variables such as property prices and the interest rates; see for example, \cite{Kau:1990aa}. Thus, under the option theoretic approach, other factors, such as the transaction costs, borrower's characteristics, etc., are assumed to have no impact on values of  the mortgage and the property underlined. The ruthless default assumption is not universally accepted in the literature and evidence against the validity of the assumption has been presented by many authors. Furthermore, as pointed out by \cite{Soyer:2010aa}, implementation of this class of models requires availability of performance level data on individual loans over time which is typically difficult to obtain. 

The alternate point of view, that does not subscribe to the ruthless default assumption, favors direct modeling of time to default of the mortgage. This approach involves hazard rate based models and also considers more direct determinants of mortgage default. This class of models includes competing risk and proportional hazards models of \cite{Lambrecht:2003aa} and duration models of \cite{Lambrecht:1997aa} that take into account individual borrower and loan characteristics. The competing risk models have been considered by many such as \cite{Deng:1996aa, Deng:2000aa}, \cite{Deng:1997aa}, and \cite{Calhoun:2002aa}. These can be considered as the competing risks versions of proportional hazards and multinomial logit models. The competing risks version of the PHM suggested by \cite{Deng:1997aa} involves evaluating hazard rates under the prepayment and default options. The author refers to these as  prepayment risk and default risk.  The competing risks approach is found to be useful in explaining the prepayment and default behaviors and improving the prediction of mortgage terminations. An application of these models to commercial mortgages can be found in \cite{Ciochetti:2002aa}.

Most of the above models use classical methods for estimation and as a result they do not provide probabilistic inferences. Some exceptions to these are the Bayesian work by \cite{Popova:2008aa} who proposed Bayesian methods for forecasting mortgage prepayment rates, \cite{Soyer:2010aa} who considered Bayesian mixtures of proportional hazards models for describing time to default and \cite{Kiefer:2010aa} who proposed an Bayesian approach for default estimation using expert information. More recently, Bayesian time series models have been considered in \cite{Aktekin:2013aa} and \cite{Lee:2016aa}. Bayesian mixture models have been considered in \cite{Klingner:2016aa}. Our approach differs from the previous in that we consider Bayesian competing risk proportional hazards models and in so doing we use both default and prepayment data. Bayesian analysis of competing risk models has been considered by \cite{Sun:1993aa} in reliability analysis and semiparametric Bayesian proportional hazards competing risk models have been introduced by \cite{Gelfand:1995aa} in survival analysis. Our work differ from these both in terms of the application and the specific approach taken.   

In this paper we consider modeling duration of single-home mortgages. In doing so, we model default and prepayment probabilities simultaneously using competing risks proportional hazards models. We include both individual and aggregate level covariates in our model. We adopt the Bayesian viewpoint in the analysis and develop posterior and predictive inferences by using Markov chain Monte Carlo (MCMC) methods. In addition to providing a formalism to incorporate prior opinion into the analysis, the Bayesian approach enables us to describe all our inferences probabilistically and provides additional insights from the analysis. In what follows, we first introduce the competing risks proportional hazards models in Section 2. The Bayesian inference is presented in Section 3 where posterior and predictive analyses are developed. In Section 4 we illustrate implementation of our model and Bayesian methods using simulated data. Concluding remarks follow in Section 5.  

\section{Competing Risk Proportional Hazards Model}
\begin{figure}[!h]
\centering
\captionsetup{justification=centering}
\includegraphics[scale=0.5]{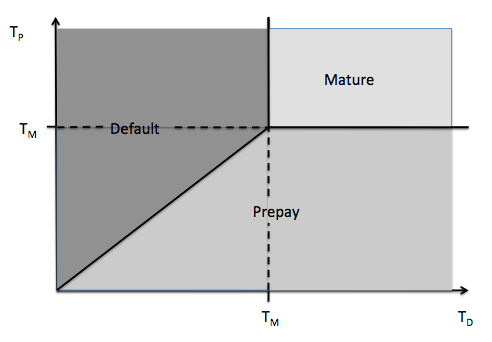}
\caption{Competing risk representation of a mortgage that can default, prepay or mature at time ${\cal T}_M$.}
\label{fig:compete}
\end{figure}

To introduce some notation let $L$ denote the mortgage lifetime and ${\cal T}_M$ denote the maturity date of the mortgage loan. Note that if a mortgage loan is not defaulted or prepaid then $L={\cal T}_M$. If we let $T_D$ and $T_P$ denote time to default and time to prepayment for a mortgage loan, respectively, then if  $(T_D>T, T_P>T)$ then $L={\cal T}_M$. Figure \ref{fig:compete} illustrates the relationship between ${\cal T}_M$, $T_D$ and $T_P$. If both $T_D$ and $T_P$ are larger than ${\cal T}_M$ then the mortgage will be paid on time. For a given mortgage loan it is of interest to infer events of ``full payment'', default and prepayment. In other words, we are interested in computing probability statements such as $P(T_D > {\cal T}_M, T_P > {\cal T}_M)$, $P(T_D < T_P \, | \, T_D < {\cal T}_M)$ or $P(L > t \, | \, L < {\cal T}_M)$.

In view of the above, we can write $$L=min(T_D, T_P, {\cal T}_M),$$ where both $T_D$ and $T_P$ are random variables. We will model $T_D$ and $T_P$ separately as  proportional hazards models (PHMs) as in Cox (1972). We denote the hazard (failure) rate for default and for prepayment as $\lambda_D(t)$ and $\lambda_P(t)$, respectively. We will refer to $\lambda_D(t)$ as the \textit{default rate} and to $\lambda_P(t)$ as the \textit{prepayment rate}. We model the default rate as 
\begin{equation} \lambda_D(t  \, | \, X(t)) = r_D(t \, | \, \psi) \, \exp(\theta_D^{\prime}X(t)), \label{eq:lambdaD} \end{equation}
where $r_D(t \, | \, \psi)$ is the baseline default rate, $\psi$ is vector of parameters, $X(t)$ is a vector of time dependent covariates and $\theta_D$ is a vector of regression parameters. Similarly, the prepayment rate is modeled as 
\begin{equation} \lambda_P(t  \, | \, X(t)) = r_P(t \, | \, \psi) \, \exp(\theta_P^{\prime}X(t)). \label{eq:lambdaP} \end{equation}
Note that the components of the covariate vector $X(t)$ may be different for the default and prepayment rates.  

We assume that default and prepayment are ``competing risks'', so that we only observe the first of them to occur. The observation of one at time $t$ implies that the other is right-censored at $t$. Thus, assuming conditional independence the joint survival function of $T_D$ and $T_P$ is given by

\begin{multline*} \mathbb{P}(T_D > t_D, T_P > t_P \, | \, r_D(), r_S(), \theta_D, \theta_P, \{ X(w) \, | \, 0 \leq w \leq \max(t_D,t_P)\}) \\ = \: \exp \left(-\int_0^{t_D}  \lambda_D(w) \: dw \: - \: \int_0^{t_P}  \lambda_P(w) \: dw\right). \end{multline*}
An active mortgage lifetime observed as $t$ implies that neither a default nor a prepayment occurs by time $t$, that is, both default and prepayment are right-censored at $t$. This includes the event that the mortgage matures at time $T$.

Empirical evidence suggests that the default rate is non-monotonic. As discussed by \citet{Soyer:2010aa}, it is reasonable to expect that the default rate is first increasing and then decreasing. A lifetime model having such hazard rate behavior is the lognormal model. Thus, we assume that baseline time to default $T_D$ follows a lognormal model with probability density function
\[ p(t_D \, | \, \mu, \sigma^2) = \frac{1}{\sqrt{2\pi \sigma^2} t_D}\exp\left(-\frac{1}{2\sigma^2}(\log (t_D) - \mu)^2 \right), \; t > 0. \]
Since it is not unreasonable to expect a similar behavior in the prepayment rate, we also assume that the baseline distribution of $T_P$ is also lognormal. Thus, the baseline model for $T_D$ will be lognormal with parameters $\mu_D$ and $\sigma_D^2$, and for prepayment with parameters $\mu_P$ and $\sigma_P^2$. 

The failure rate of the lognormal distribution can be written in terms of the standard Gaussian distribution function $\Phi$. In fact the failure rates for $T_D$ and $T_P$ then take the form:
\begin{equation} 
\lambda_D(t  \, | \, X(t)) = \frac{(2\pi \sigma_D^2)^{-1/2} t^{-1} \exp\left(-0.5(\log (t) - \mu_D)^2 / \sigma_D^2 \right)}{1 - \Phi((\log (t) - \mu_D)/\sigma_D)} \: \exp\left(\theta_D^{\prime} X(t) \right)
\label{eq:failurerateD}
\end{equation}
and
\begin{equation} 
\lambda_P(t  \, | \, X(t)) = \frac{(2\pi \sigma_P^2)^{-1/2} t^{-1} \exp\left(-0.5(\log (t) - \mu_P)^2 / \sigma_P^2 \right)}{1 - \Phi((\log (t) - \mu_P)/\sigma_P)} \: \exp\left(\theta_P^{\prime} X(t) \right);
\label{eq:failurerateP}
\end{equation}
see the Appendix for details on derivation of \ref{eq:failurerateD} and \ref{eq:failurerateP}.

\section{Bayesian analysis of the competing risk PHM} 
\label{sec:nonhetero}

We assume that data on $N$ mortgages are available. From these, $n_D$ have defaulted, $n_P$ have prepayed and $n_C = N - n_D - n_P$ are still active, including those that have matured successfully. The $N$ mortgages are indexed $i = 1,\ldots,n_D$ for the defaulted mortgages, $i=n_D+1,\ldots,n_D + n_P$ for prepaid and $i = n_D + n_P + 1,\ldots,N$ for active. 
Let $\bm{t_D} = \{t_1^D,\ldots,t_{n_D}^D \}$ be the times of default and $\bm{t_P} = \{t_{n_D+1}^P,\ldots,t_{n_D+n_P}^P \}$ be the times of prepayment. For the $n_C$ mortgages that are still active, let $\bm{t_C} = \{t_{n_D+n_P+1}^C,\ldots,t_{N}^C \}$ be the times since the initiation of mortgages; $t_i^C = {\cal T}_M$ for those that have matured.  

Also observed are the covariates. Let $\bm{X}_i(t) = (X_{i1}(t),\ldots,X_{im}(t))$ be the vector of covariates for mortgage $i$ at time $t$.  Some of these are common covariates e.g.\ interest rates, while others are mortgage specific e.g.\ mortgage size or credit score. We assume that they are observed at a known set of times $\tau_1 <\tau_2 <\cdots <\tau_m$ and that they are piecewise constant on intervals for which these times are the mid-points. Hence
\begin{equation} 
\bm{X}_i(t) = \bm{X}_i(\tau_j), \; s_{j-1} < t \leq s_j, 
\label{eq:piecewise}
\end{equation}
for $j=1,\ldots,m$, where $s_0 =0$, $s_j = 0.5(\tau_j + \tau_{j+1})$ for $j=1,\ldots,m-1$ and $s_m = \infty$.
 We let $\bm{X}_i = \{ \bm{X}_i(\tau_1),\ldots,\bm{X}_i(\tau_m) \}$ be the observed covariates for mortgage $i$ and $\bm{{\cal X}} = \{ \bm{X}_i(\tau_k) \, | \, i=1,\ldots,N; \, k=1,\ldots,m \}$ be the set of all observed covariates. 

%\subsection{Derivation of the Posterior Distribution}
The unknown quantities in this model are the regression parameters $\theta_D$ and $\theta_P$, and the baseline failure rate parameters $\psi = (\mu_D,\sigma_D^2,\mu_P,\sigma_P^2)$. The required posterior distribution is therefore:
\begin{equation}
p(\theta_D,\theta_P, \psi \, | \, \bm{t_D}, \bm{t_P}, \bm{t_C}, \bm{{\cal X}} )
\: \propto \: p(\bm{t_D}, \bm{t_P}, \bm{t_C} \, | \, \theta_D,\theta_P, \psi, \bm{{\cal X}}) p(\theta_D) p(\theta_P) p(\psi).% \: p({\cal X}_{-}(t) \, | \, \bm{X},\psi_X) \: p(\theta_D,\theta_P,\psi_X).
\label{eq:posterior}
\end{equation}
For the likelihood term $P(\bm{t_D}, \bm{t_P}, \bm{t_C} \, | \, \theta_D,\theta_P, \psi,\bm{{\cal X}})$, we assume observations are conditionally independent, given the parameters. From the competing risks assumption, an observation $t_i^D$ is an exact observation of $T_D$ and a right-censored observation of $T_P$; it is vice versa for $t_i^P$. Finally, $t_i^C$ is a right-censored observation of both $T_D$ and $T_P$. Hence:
\begin{multline} 
p(\bm{t_D}, \bm{t_P}, \bm{t_C} \, | \, \theta_D,\theta_P,  \psi,\bm{{\cal X}}) \\
\: = \: \left( \prod_{i=1}^{n_D} p(t_i^D \, | \, \theta_D, \bm{X}_i) \: P(T_P > t_i^D \, | \, \theta_P,  \bm{X}_i) \right) \; \left( \prod_{i=n_D+1}^{n_D+n_P} p(t_i^P \, | \, \theta_P, \bm{X}_i) \: P(T_D > t_i^P \, | \, \theta_D,  \bm{X}_i) \right) \\ \times \;  \left( \prod_{i=n_D + n_P + 1}^{N} P(T_D > t_i^C \, | \, \theta_D, \bm{X}_i) \: P(T_P > t_i^C \, | \, \theta_P, \bm{X}_i) \right) \\
\!\!\!\!\!\!\!\!\!\!\!\!\!\!\!\!\!\!\!\!\!  = \: \left( \prod_{i=1}^{n_D} \lambda_D(t_i^D \, | \, \bm{X}_i(t_i^D)) \exp\left(-\int_0^{t_i^D} \lambda_D(w\, | \, \bm{X}_i(w)) + \lambda_P(w \, | \, \bm{X}_i(w)) \: dw\right) \right) \\ \times \;  \left( \prod_{i=n_D+1}^{n_D+n_P} \lambda_P(t_i^P \, | \, \bm{X}_i(t_i^P)) \exp\left(-\int_0^{t_i^P} \lambda_D(w \, | \, \bm{X}_i(w)) + \lambda_P(w \, | \, \bm{X}_i(w)) \: dw\right) \right) \\ \times \;  \left( \prod_{i=n_D + n_P + 1}^{N} \exp\left(-\int_0^{t_i^C} \lambda_D(w \, | \, \bm{X}_i(w)) + \lambda_P(w \, | \, \bm{X}_i(w)) \: dw\right) \right),
\label{eq:likelihood}
\end{multline}
where $\lambda_D(t \, | \, \bm{X}_i(t))$ and $\lambda_P(t \, | \, \bm{X}_i(t))$ are given by Equations \ref{eq:failurerateD} and \ref{eq:failurerateP}, $\bm{X}_i(t)$ is given in Equation \ref{eq:piecewise} and a formula for the integrals is given in Equation \ref{eq:lambdaD_integral2} of the Appendix. The formula for the integrals becomes tricky in practice for time varying covariates, as noted in Cox and Oakes (1984).

An independent zero-mean normal prior is assumed for each component of $\theta_D$ and $\theta_P$, as well as $\mu_D$ and $\mu_P$. 
%{\color{blue} For $\sigma_D$ and $\sigma_P$, since no prior provides us with known full conditionals, an exponential prior is assumed. It is noted that a lognormal or gamma prior could also have been chosen.} 

The model above is such that the parameters are not identifiable without further assumptions. For a Bayesian analysis, such as ours, that means whether the data can inform well enough about all the parameters. Identifiability issues can be overcome via specific prior specification or model dimension reduction \citep{Gelfand:1995aa}. We will see in the results that for the default model, identifiability is present. It is attributable to the low number of default mortgages under which parameter learning becomes very difficult; no such issue exists for the parameters under the prepaid model.

%\subsection{Computation of the Posterior by MCMC}
An MCMC procedure, based on the Metropolis within Gibb's sampler \citep{Tierney:1994aa}, has been implemented to sample from $p(\theta_D,\theta_P, \psi \, | \, \bm{t_D}, \bm{t_P}, \bm{t_C}, \bm{{\cal X}})$. The covariate coefficient vectors $\theta_D$ and $\theta_P$ are updated as blocks from their full conditional distributions with a Gaussian random walk proposal, while each component of $\psi$ is updated separately.  The Appendix contains the details of the algorithm.

The MCMC output is a set of samples of all the unknowns from the posterior distribution. Let the number of samples be $L$, and let $\theta_D^{(l)}$, $\theta_P^{(l)}$ and $\psi^{(l)}$ denote the $l$th samples of $\psi$, $\theta_D$ and $\theta_P$ respectively.

%\subsection{Prediction}
%\label{subsec:pred}
The MCMC output can be used to compute many quantities of interest. 
%Many of these are probabilistic predictions or expectations that are based on the posterior distribution. These are approximated by the MCMC samples through the principle of Monte Carlo integration (an application of the Law of Large Numbers).
With the posterior samples, one can compute for a mortgage with a known set of covariates $\bm{X}=\{\bm{X}(w) \,| \, w \geq 0 \}$:
%\begin{itemize}
The posterior predictive reliability function of the time to default is approximated by
\begin{equation}
P(T_D > t \, | \, \bm{t_D}, \bm{t_P}, \bm{t_C}, \bm{{\cal X}}, \bm{X}) \: \approx \: \frac{1}{G} \sum_{l=1}^G \exp\left( - \int_0^{t} \lambda_D^{(l)}(w) \: dw\right), 
\label{eq:post_default}
\end{equation}
and the time to prepayment is approximated by
\begin{equation}
P(T_P > t \, | \, \bm{t_D}, \bm{t_P}, \bm{t_C}, \bm{{\cal X}}, \bm{X}) \: \approx \: \frac{1}{G} \sum_{l=1}^G \exp\left( - \int_0^{t_P} \lambda_P^{(l)}(w) \: dw\right),
\label{eq:post_prepay}
\end{equation}
where $\lambda_D^{(l)}(w) = r_D^{(l)}(w \, | \, \psi) \exp(\theta_D^{(l) \, \prime}\bm{X}(w))$ and $\lambda_P^{(l)}(w) = r_P^{(l)}(w \, | \, \psi) \exp(\theta_P^{(l) \, \prime}\bm{X}(w))$, the values of $r_D^{(l)}(w \, | \, \psi)$ and $r_P^{(l)}(w \, | \, \psi)$ are given by Equation \ref{eq:failurerate}, using the parameter values in $\psi^{(l)}$, and a formula for the integrals is given by Equation \ref{eq:lambdaD_integral2} of the Appendix. 

Equations \ref{eq:post_default} and \ref{eq:post_prepay} allows us to determine, by simulation, the probability that a mortgage will default, prepay or mature with a given set of covariates $\bm{X}$.  The inverse distribution method can be used to simulate independently many values pairs $(t_D,t_P)$ from these reliability functions e.g.\ for $t_D$, generate a random number $u$ and then solve $u = P(T_D > t \, | \, \bm{t_D}, \bm{t_P}, \bm{t_C}, \bm{{\cal X}}, \bm{X})$ for $t$, an easy numerical exercise. This further means we can compute predictive densities $P(T_D \, | \, \bm{t_D}, \bm{t_P}, \bm{t_C}, \bm{{\cal X}}, \bm{X})$ and $P(T_P \, | \, \bm{t_D}, \bm{t_P}, \bm{t_C}, \bm{{\cal X}}, \bm{X})$ for each mortgage as well. Furthermore to this, the probabilities that a loan defaults, prepays or matures are approximated by the proportion of simulated pairs $(t_D,t_P)$ that lie in their respective regions as defined in Figure \ref{fig:compete}:
\begin{eqnarray*}
\mbox{Defaults}  & \Leftrightarrow & t_D   <  {\cal T}_M \mbox{ and } t_D < t_P; \\
\mbox{Prepays}   & \Leftrightarrow & t_P   <  {\cal T}_M \mbox{ and } t_P < t_D; \\
\mbox{Matures}   & \Leftrightarrow & t_D \geq {\cal T}_M \mbox{ and } t_P \geq {\cal T}_M.
\end{eqnarray*}
%\end{itemize}
Since the marginal density is
\[ \displaystyle f(t) = \lambda(t)\, R(t),\]
where $\lambda(t)$ is failure rate and $R(t)$ is reliability function as defined in equations \ref{eq:post_default} and \ref{eq:post_prepay}. So given $R(t)$ for $T_D$ we can compute the pdf of $T_D$
\[ \displaystyle f_D(t) \: \approx \: \frac{1}{G} \sum_{l=1}^G \lambda_D^{(l)}(t) \exp\left( - \int_0^{t} \lambda_D^{(l)}(w) \: dw\right).\]
Similarly compute for pdf of $T_P$.

\section{The Freddie Mac Single Family Loan Dataset}

The Federal Home Loan Mortgage Corporation (FHLMC), known as Freddie Mac, is a public company that is sponsored by the United States government. It was formed in 1970 to expand the secondary market for mortgages in the US. It has provided a dataset about single family loan-level credit performance data on a portion of fully amortizing fixed-rate mortgages that the company purchased or guaranteed. The dataset contains information about approximately 21.5 million fixed-rate mortgages that originated between January 1, 1999, and December 31, 2014. The dataset can be downloaded from the Freddie Mac website and is organised as two files for each quarter:
\begin{enumerate}
\item the origination data file that contains data concerning the set up of the loan;
\item the monthly performance data file that contains the monthly performance of each loan e.g.\ amount repaid, the outstanding principal, whether it is in default, etc. 
%{\color{red} @ARNAB: so there is one of these files for each month, not for each quarter?} {\color{blue} @SIMON: No there is a single file per quarter. The year and quarter refers to loan origination date.}
\end{enumerate}
There is also a smaller sample data set that contains a simple random sample of 50,000 loans selected from each year and a proportionate number of loans from subsequent years  (the actual definition is 50,000 loans selected from each full vintage year and a
proportionate number of loans from each partial vintage year of the full single family
loan-level data set). 
%{\color{red} @ARNAB: not sure what the proportionate bit means - is the data not from 15 full years?} {\color{blue} @SIMON: this has been explained by Miss Song. It is worth noting that the data set is "live", i.e.\ it gets refreshed every now and then. So there could be new loans which have originated in $2016$, which is the partial vintage year. It may not be possible to select $50000$ loans from partial vintage year. So they just sample a proportionate number from $2016$.}. 
The sample data set also has an origination and monthly performance file for each year 
%{\color{red} @ARNAB: per year or per quarter?}  {\color{blue} @SIMON: per year.}.

Some processing of the raw data was needed to transform it into a format that can be be analysed by this model.  Each loan was tracked through the data to categorize it as active, defaulted or prepaid. Since loans in the dataset originated in 1999 and were for 30 years, there were no loans classified as mature and so this category could be ignored. 

\subsection{Loan categorization}
Four fields in the data were used to categorize each loan as default, prepay or active, and to define the observed time:
\begin{itemize}
\item \texttt{zero\char`_balance} defines whether a particular loan's balance has reduced to 0 or
not, and has the following codes:
\begin{description}
\item[01] Prepaid or Matured (voluntary payoff);
\item[03] Foreclosure Alternative Group (Short Sale, Third Party Sale, Charge Off or Note Sale);
\item[06] Repurchase prior to Property Disposition;
\item[09] REO Disposition; and
\item[empty] Not Applicable.
\end{description}
\item \texttt{delinquency} provides a value corresponding to the number of days the borrower has not paid the loan, according to the due date of last paid installment, or if a loan is acquired by REO, coded as:
\begin{description}
\item[0] Current, or less than 30 days past due;
\item[1] 30--59 days delinquent;
\item[2] 50--89 days delinquent;
\item[3] 90--119 days delinquent, etc.;
\item[R] REO acquisition;
\item[empty] Unavailable.
\end{description}
\item \texttt{reporting\char`_date} is the month that the observation is made in.
\item \texttt{months\char`_remain} is the number of months until the legal maturity of the loan.
\end{itemize}
Then the loan status was defined as:
\begin{itemize}
\item \textbf{Prepaid} if there exists a month where \texttt{zero\char`_balance} = 01 AND  \texttt{repurchase} = ``N". In this case, the prepaid time $t_P$ is the time from loan origination to the \texttt{reporting\char`_date} where this first happens. 
%{\color{blue} @SIMON: I have corrected this. As per new categorization, a mortgage is prepaid if \texttt{zero\char`_balance} = 01 AND  \texttt{repurchase} = ``N".}
%\item \textbf{Prepaid} if there exists a month where \texttt{zero\char`_balance} = 01 AND  \texttt{months\char`_remain} $>=$ 1. In this case, the prepaid time $t_P$ is the time from loan origination to the \texttt{reporting\char`_date} where this first happens.
\item \textbf{Default} if there exists a month where \texttt{zero\char`_balance} = 03, 06 or 09.  In this case, the default time $t_D$ is the time from loan origination to the \texttt{reporting\char`_date} where this first happens.
% \item \textbf{Matured} if there exists a month where \texttt{zero\char`_balance} = 01 AND  \texttt{months\char`_remain} $=$ 0.  For this analysis, the loan is considered as active for the time from loan origination to the \texttt{reporting\char`_date} where this happens.
\item \textbf{Active} if the loan could not be classified as Prepaid or Active AND the latest \texttt{reporting\char`_date} corresponding to the loan is later than 01/01/2014 AND \texttt{zero\char`_balance} is empty at that latest date AND \texttt{delinquency} is not equal to R at that latest date. The active time is the time from loan origination to the \texttt{reporting\char`_date} where this happens.
\end{itemize}
These definitions are not exhaustive; there are loans in the dataset that are discontinued without any clear information and such loans have been excluded from our analysis.

\subsection{Covariates}
The following covariates (fixed term) are available in the dataset: credit score, mortgage insurance percentage (MI), number of units, combined loan-to-value (CLTV), debt-to-income (DTI), unpaid principal balance (UPB), original interest rate, number of borrowers, first homebuyer, occupancy status, property type, property state (state in which property resides) and current interest rate. 
%{\color{blue} Original interest rate refers to the rate at loan initiation while current interest rate contains monthly interest rates since loan start time. So the latter is the only variable whose values changes with time. Out of the rest,} 
first homebuyer, occupancy status, property type and property state are categorical variables and have been converted to indicator variables. The covariate property state has been re-categorized into judicial or non-judicial state. For the rest of the categorical variables, some of categories were of low frequency. For example, there are 6 categories in variable \texttt{property\char`_type}, of which $81\%$ were \texttt{single family home} and some categories like \texttt{leasehold} accounting for as low as $0.0003\%$. It was decided to group categories with extremely low frequencies for all the categorical variables. All the quantitative variables have been standardized. Furthermore strong correlation have been found between \texttt{mortgage insurance percentage} and \texttt{combined loan-to-value}, and between original and current \texttt{interest rates} which led us to drop the latter in both the cases. Since \texttt{current interest rate} has been dropped we do not have to work with any time dependent covariate.
%{\color{red} @ARNAB: assuming that first time home-buyer flag is 1 or 0, number of units is positive integer.  What is the range of values of the other covariates? What is property state?}  {\color{blue} I have updated the paragraph. See if this is alright.}

%The following covariates (fixed term) are available in the dataset: credit score, mortgage insurance percentage, number of units, combined loan-to-value, debt-to-income, UPB, original insurance rate, number of borrowers, first homebuyer, occupancy status, property type, property state (state in which property resides) and current interest rate. Among these covariates first homebuyer, occupancy status, property type and property state are categorical variables and have been converted to indicator variables. The covariate property state has been re-categorized into judicial or non-judicial state. Apart from these, we have current interest rate as the only real time covariate. {\color{red} @ARNAB: assuming that first time home-buyer flag is 1 or 0, number of units is positive integer.  What is the range of values of the other covariates? What is property state?}  {\color{blue} Do you want me list out some of the categories? What about continuous covariates?}
\section{Analysis of the Data}
\begin{figure}[!h]
\centering
\captionsetup{justification=centering}
\includegraphics{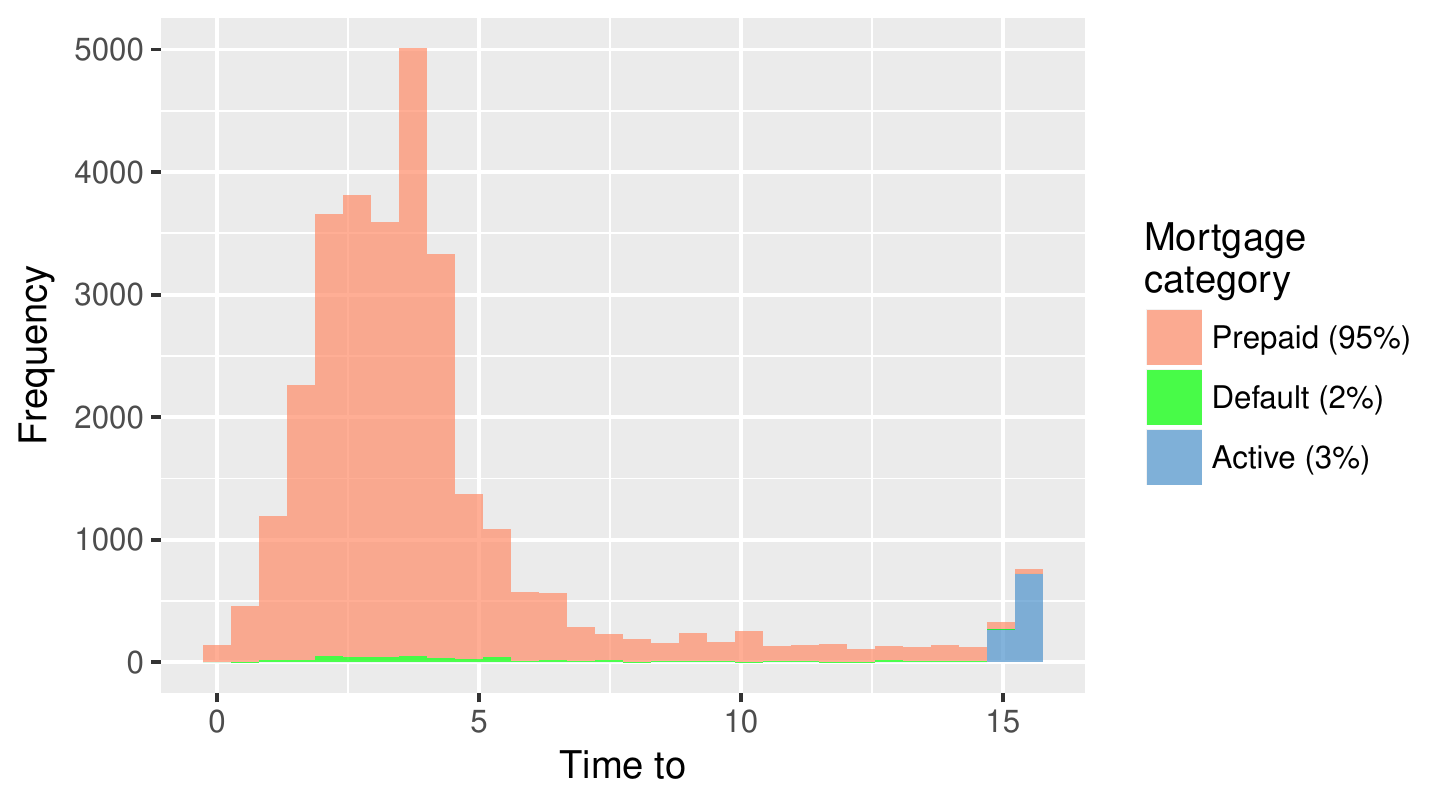}
\caption{Histogram of time to default, prepaid and active times for each category.}
\label{fig:mortgage_cat_hist}
\end{figure}

The data set comprised of $672208$ mortgages originating in the year 1999. This data set is extremely unbalanced with $95\%$ of the mortgages being \texttt{prepaid}, $3\%$ being \texttt{active} and the only about $1.6\%$ belonging to \texttt{default} category. This huge imbalance is evident in figure \ref{fig:mortgage_cat_hist}.
%\begin{centering}
%\end{centering}

%\begin{centering}
%\begin{figure}[!h]
%\includegraphics{time_to_box}
%\caption{Boxplot of time to default, prepaid and active times for each category.}
%\label{fig:mortgage_cat_box}
%\end{figure}
%\end{centering}

%{\color{blue} Both $\theta_D$ and $\theta_P$ had Normal prior with zero mean and standard deviation $100$. The mean parameters $\mu_D$ and $\mu_P$ have zero mean Normal priors with standard deviation $10$, while an exponential prior with mean $100$ was assumed for the standard deviation parameters $\sigma_D$ and $\sigma_P$. The starting value of each of the parameter chains were randomly selected using Normal and inverse gamma distributions for mean and standard deviation parameters respectively. The standard deviation for the proposal distributions, for example $s_{\theta, D}^2$ or $s_{\mu, D}^2$ have also been generated randomly from inverse gamma distributions to provide more diversity in the chains. The scale parameter ($a$) used in proposal for $\sigma_D$ and $\sigma_P$ was generated from uniform distribution.}
 
\begin{figure}[!h]
\centering
\captionsetup{justification=raggedright}
\includegraphics{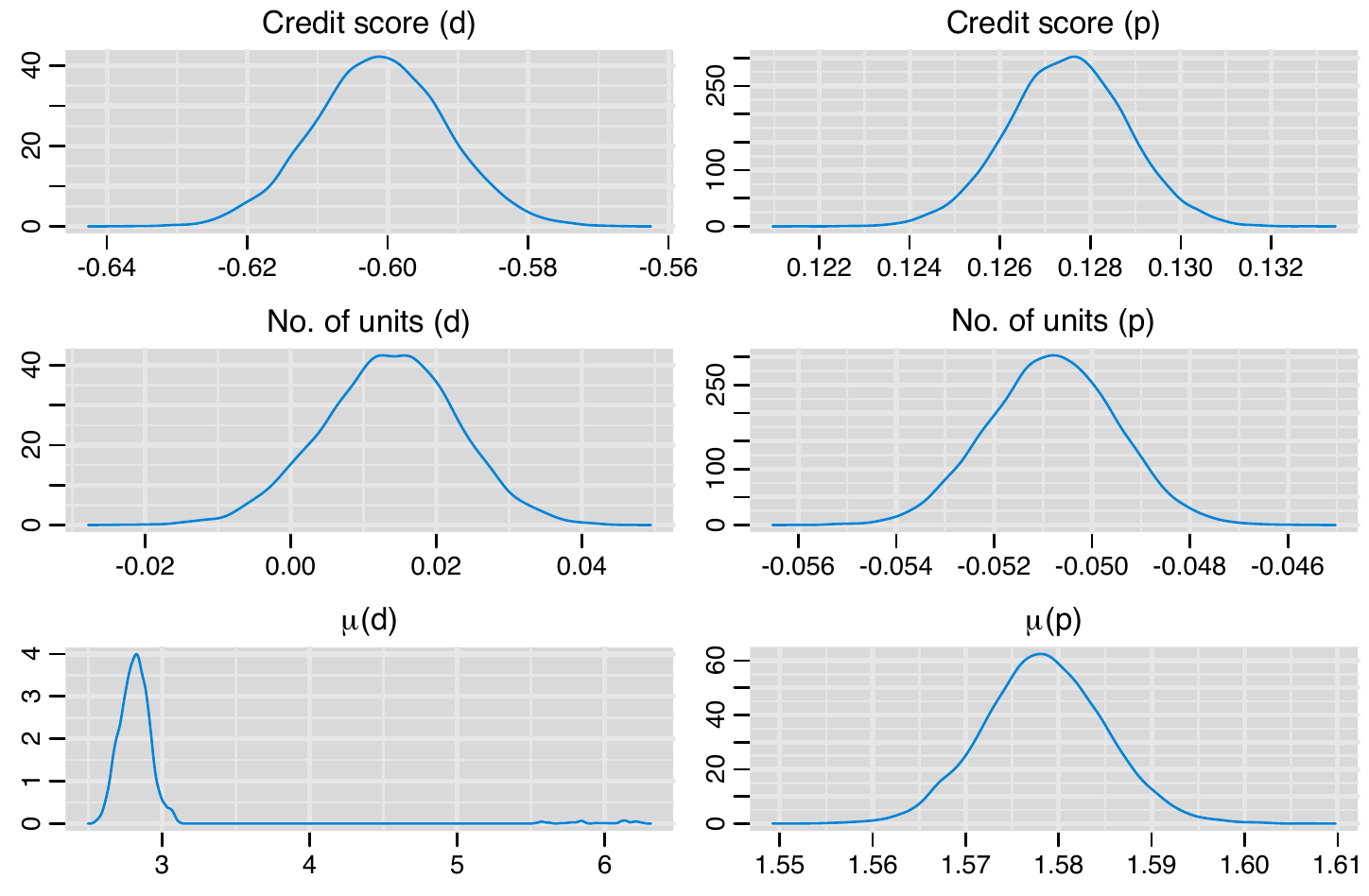}
\caption{Density plot of combined samples from merging all the chains for credit score, number of units and $\mu$, both for default (d) and prepaid (p) times. The long tail corresponding to $\mu$(d) can be attributed to a single chain which is slow in converging.}
\label{fig:mcmc_subset}
\end{figure}

Rcpp \citep{Eddelbuettel:2011aa} has been used to construct the MCMC algorithm. This has greatly improved the speed of the algorithm given that the data set is extremely large. The MCMC procedure was run in $50$ chains for $75,000$ iterations each. We set burnin at $60000$ and thinned the remaining by selecting every $50^{th}$ sample. Trace plots, provided in the Appendix, for all the parameters show good mixing for all the covariates implying convergence. We provide the density plot constructed by combining the thinned chains for a subset of covariates in figure \ref{fig:mcmc_subset}. 
%{\color{blue}The skew in the density of $\mu_D$ is caused by the slow convergence of a single chain. We assume that this single chain suffers from slow convergence due to the problem of identifiability in proportional hazards models. The problem is further enhanced by the fact that the number of default mortgages is very low, hence making sufficient learning difficult.}

%\begin{centering}
%\end{centering}

Tables \ref{tab:posterior_summary_baseline} and \ref{tab:posterior_summary_covar} are summaries of the marginal posterior distributions of the model parameters, based on the 15,000 combined samples of the MCMC.  We see in table \ref{tab:posterior_summary_covar} that nearly all the covariates, with the exception of \texttt{no. of units}, turn out to be significant. \texttt{Credit score}, \texttt{UPB}, \texttt{no. of units}, \texttt{type of property} and \texttt{no. of borrowers} have opposite effects on default and prepay rates. Default rate is found to decrease with \texttt{credit score}, \texttt{UPB} etc, as it should be, and prepay rate increases for the same. Other variables, for example, \texttt{DTI}, \texttt{mortgage insurance \%}, \texttt{original interest rate}, \texttt{first time homebuyer}, \texttt{occupancy status} and \texttt{property state} have same signs of coefficients for both of default and prepay. Thus we can see that for \texttt{mortgage insurance \%} both default and prepay rate increase, whereas for \texttt{first time homebuyer} both the rates decrease. Also note that the estimated mean parameter (as also for the standard deviation parameter) of the baseline default rate is substantially greater than that of prepay.
%The caterpillar plot (\ref{fig:cater_allpar}) provided in the appendix gives a visual representation of the summary statistics provided in these tables.

\begin{table}
\centering
\begin{tabular}{ccc} \hline
\textbf{Parameter} & Median & 95\% Prob.\ Interval \\ \hline
$\mu_D$ & 2.817 & (2.631, 3.077) \\
$\sigma_D$ & 0.963 & (0.916, 1.028) \\
$\mu_P$  & 1.578 & (1.566, 1.591) \\
$\sigma_P$  & 0.717 & (0.713, 0.721) \\ \hline
\end{tabular}
\vspace{2mm}
\caption{\label{tab:posterior_summary_baseline}Summary of the marginal posterior distributions of the baseline default and prepay rates.}
\end{table}      

\begin{table}
\centering
\begin{tabular}{lccc|ccc} \hline
\textbf{Covariate}            & \multicolumn{2}{c}{\textbf{Default}} & \multicolumn{2}{c}{\textbf{Prepay}} \\
                              & Median & 95\% Prob.\ Interval     & Mean & 95\% Prob.\ Interval    \\ \hline
Credit score                  & -0.601 & (-0.620, -0.583)  & 0.128 & (0.125, 0.130)  \\
Mortgage insurance \%   & 0.395 & (0.376, 0.415)  & 0.068 & (0.065, 0.070)  \\
Number of units               & 0.014 & (-0.005, 0.031)  & -0.051 & (-0.053, -0.048)  \\
%Original combined loan-to-value      & 0.124 & (0.5, 0.98) & 0.01 & (-0.009, 0.03) \\
Original DTI    & 0.124 & (0.103, 0.146)  & 0.020 & (0.018, 0.023)  \\
UPB                              & -0.069 & (-0.093, -0.046)  & 0.305 & (0.302, 0.307)  \\
Original interest rate             & 0.412 & (0.396, 0.429)  & 0.376 & (0.374, 0.379)  \\
No. of borrowers            & -0.296 & (-0.316, -0.276)  & 0.055 & (0.052, 0.058)  \\
Intercept            & -3.090 & (-3.356, -2.694)  & 0.182 & (0.158, 0.207)  \\
First time home-buyer         & -0.244 & (-0.293, -0.194)  & -0.009 & (-0.016, -0.003)  \\
Occupancy status              & 0.460 & (0.342, 0.575)  & 0.249 & (0.237, 0.261)  \\
Property state                & -0.110 & (-0.149, -0.071)  & -0.080 & (-0.085, -0.075)  \\
Property type        & 0.304 & (0.249, 0.362)  & -0.061 & (-0.67, -0.054)  \\ \hline
%Current interest rate        & 0.18 & (-0.15, 0.61) & 0.25 & (-0.29, 0.80) \\ 
\end{tabular}
\vspace{2mm}
\caption{\label{tab:posterior_summary_covar} Summary of the marginal posterior distributions of $\theta_D$ and $\theta_P$.  The $95\%$ probability intervals are the $2.5 - 97.5$ percentiles of the sampled parameter values. Number of units under default mortgages is the only covariate which can be termed to be not-significant, since the CI does not contain 0.}
\end{table}

%{\color{blue} The posterior predictive densities for time to default (or prepay) corresponding to each mortgage can be computed using Equations \ref{eq:post_default} and \ref{eq:post_prepay} (see discussion following these two equations). Figure \ref{fig:pred_distr} shows predictive posterior distribution of time to default for 3 randomly chosen mortgages in the top panel. A similar plot corresponding to time to prepay has been provided in the lower panel. The posterior predictive densities for time to default are generally found to be flat, while those for prepay have variable shapes. Note that the area under the densities do not necessarily sum to 1 in these plots since we have truncated them at 2029, the maturity date of the mortgages.}
 \begin{figure}[!h]
\centering
\captionsetup{justification=raggedright}
\includegraphics{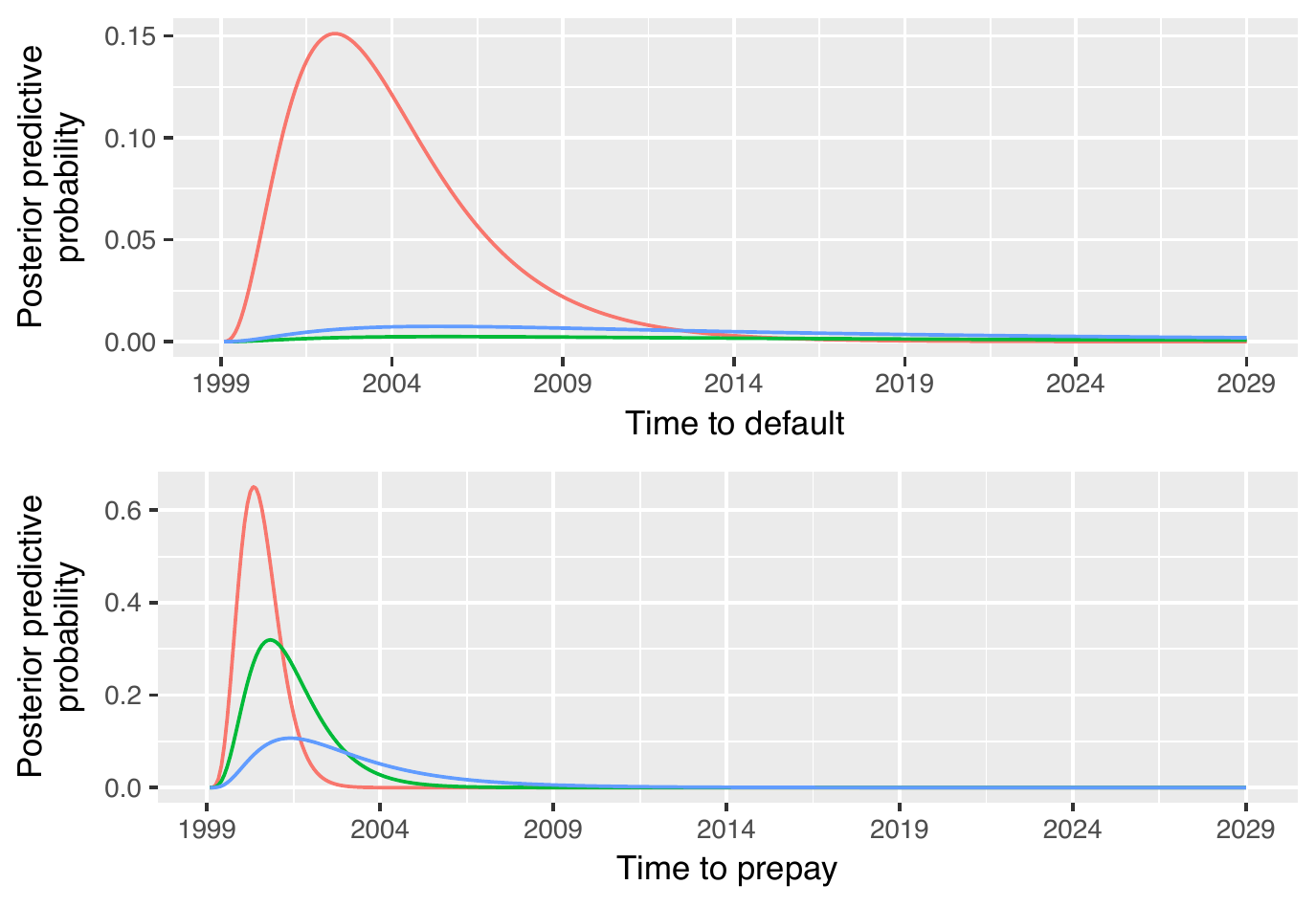}
\caption{The two panels provide posterior predictive distributions of time to default and prepay respectively. The flat posterior predictive distribution of time to default is very common in almost all mortgages, while for prepay the shape of the distributions are quite varying. Note that some of the distributions are truncated at $2029$, the year the mortgages end.}
\label{fig:pred_distr}
\end{figure}

The mortgages and their covariate values are provided in table \ref{tab:select_mortgages}
\begin{table}
\centering
\begin{tabular}{lccc|ccc} \hline
\textbf{Covariate}            & \multicolumn{3}{c}{\textbf{Default}} & \multicolumn{3}{c}{\textbf{Prepay}} \\
Mortgage number & 1 & 2 & 3 & 1 & 2 & 3  \\ \hline
Credit score                  & 724 & 541 & 750 & 787 & 668 & 619  \\
Mortgage insurance \%   & 12 & 30 & 0  & 0 & 0 & 30  \\
Number of units               & 1 & 1 & 1  & 1 & 1 & 1  \\
Original DTI    & 16 & 27 & 23  & 39 & 34 & 44  \\
UPB                              & 73000 & 112000 & 83000  & 37000 & 312000 & 204000  \\
Original interest rate             & 6.875 & 10 & 8  & 6.875 & 7 & 9.625  \\
No. of borrowers            & 2 & 2 & 2  & 1 & 2 & 2  \\
First time home-buyer         & No & No & No & No & No & No  \\
Occupancy status              & Owner & Owner & Owner & Owner & Owner & Owner  \\
Property state                & Non-Jud & Jud & Non-Jud  & Non-Jud & Non-Jud & Non-Jud  \\
Property type        & SF  & SF & SF  & SF & SF & SF  \\ \hline
\end{tabular}
\vspace{2mm}
\caption{\label{tab:select_mortgages} The values of the covariates for the 6 mortgages that have been used for computing the posterior predictive densities in figure \ref{fig:pred_distr} is provided here. Abbreviations used are ``Owner" - ``Owner occupied", ``Non-Jud"/``Jud" - ``Non-Judicial"/``Judicial" and ``SF" - ``Single family".}
\end{table}
    
\section{Model Assessment}
The suitability of the model is assessed by deriving, for each loan in the data:
\begin{itemize}
\item The probabilities that the loan defaults, prepays or remains active up to the end of the data, following the method in Section \ref{sec:nonhetero}, which can be compared to the actual outcome; 
\item If the mortgage defaulted then the predicted reliability function of the default time can also be computed from Equation \ref{eq:post_default}, and hence the quantile of the observed time.  A standardised residual can also be computed e.g.\ $(t_D - E(t_D))/sd(t_D)$, where $t_D$ is the observed default time, $E(t_D)$ and $sd(t_D)$ are the mean and standard deviation of the posterior default time, derived from the predicted reliability function.
\item Similarly, if the mortgage was prepaid then the predicted reliability function of the prepay time can be computed from Equation \ref{eq:post_prepay}. The quantile of the observed prepay time and a standardised residual can be derived.
\end{itemize}
Active loans are right-censored observations of both the default and prepay times.  The competing hazards model implies that default times are also right-censored observations of a prepay time, and vice versa.        

\begin{figure}[!h]
\centering
\captionsetup{justification=centering}
\includegraphics{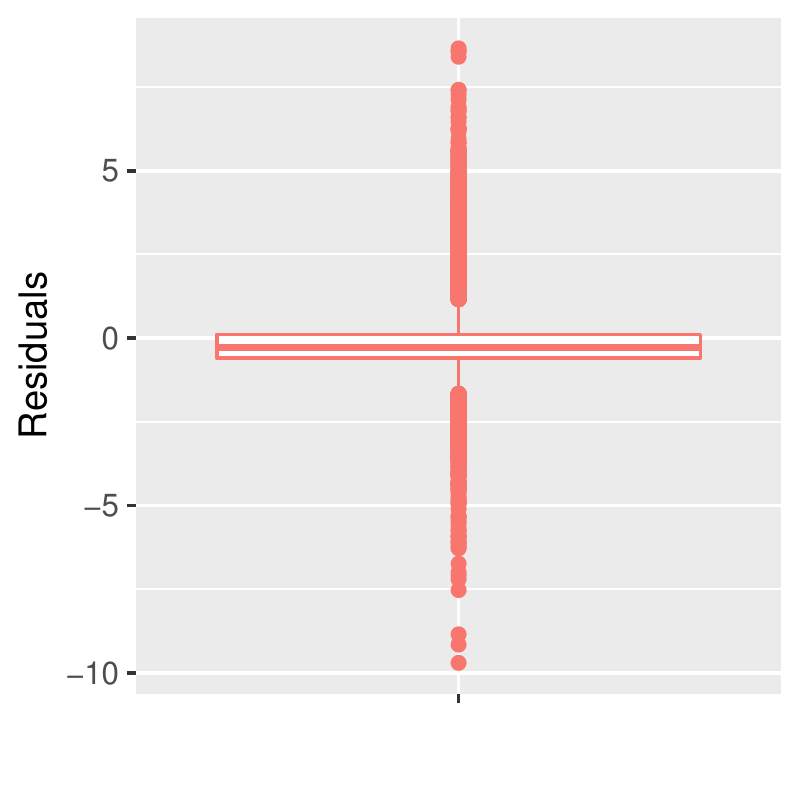}
\caption{Residuals of all the default and prepay mortgages combined. Note the slight bias below $0$ which is caused by the default mortgages.}
\label{fig:residual}
\end{figure}

We assessed the fitted model on the sample data set in the year 1999, which has $30755$ mortgages. Figure \ref{fig:residual} shows a box plot of standardised residuals (as explained above) for all the default and the prepaid mortgages and is found to be centered around 0. If we isolate the defaulted mortgages we find that the corresponding residuals are biased away from 0. Identifying mortgages that defaulted is found to be difficult form that data we have since they constitute less than $2\%$ of the whole set.

%{\color{blue} Separate box plots of standardised residuals for mortgages show a good fit for prepay, where residuals a slightly biased away from 0. However, for default mortgages almost all residuals are negative, which shows that the estimated mean time to default is over-biased.  We think that this can be largely attributed to huge contrast in proportion of each type of mortgages in the data set.}

\begin{figure}[!h]
\centering
\captionsetup{justification=centering}
\includegraphics{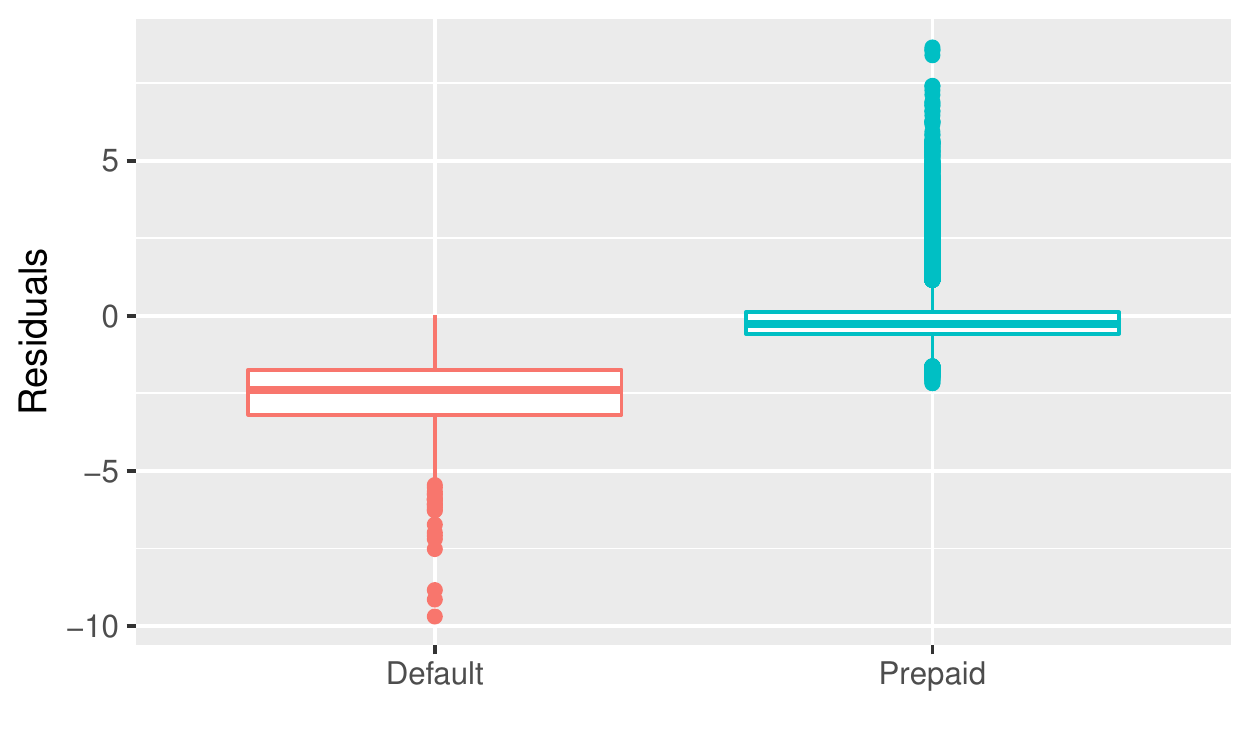}
\caption{Separate box plots of residuals corresponding to default and prepay mortgages. The residuals for default mortgages show clear bias below 0.}
\label{fig:residual2}
\end{figure}
The model was able to correctly identify approximately $50\%$ of the default mortgages and $95\%$ of the prepaid ones using $95\%$ prediction interval. The contrast between default and prepaid mortgages is also evident when we calculated the predicted reliability function for each. The median predicted reliability function for default mortgages is found to be $0.976$ and $(2.5, 97.5)$ quantiles being $(0.740, 0.999)$, while those for prepaid are $0.524\, (0.041, 0.981)$. 
%{\color{blue} The box plots in figure \ref{fig:box_surv} demonstrates the range of posterior reliability corresponding to both default and prepaid mortgages. Reliability is computed at the time to default or prepay.}
\begin{figure}[!h]
\centering
\captionsetup{justification=centering}
\includegraphics{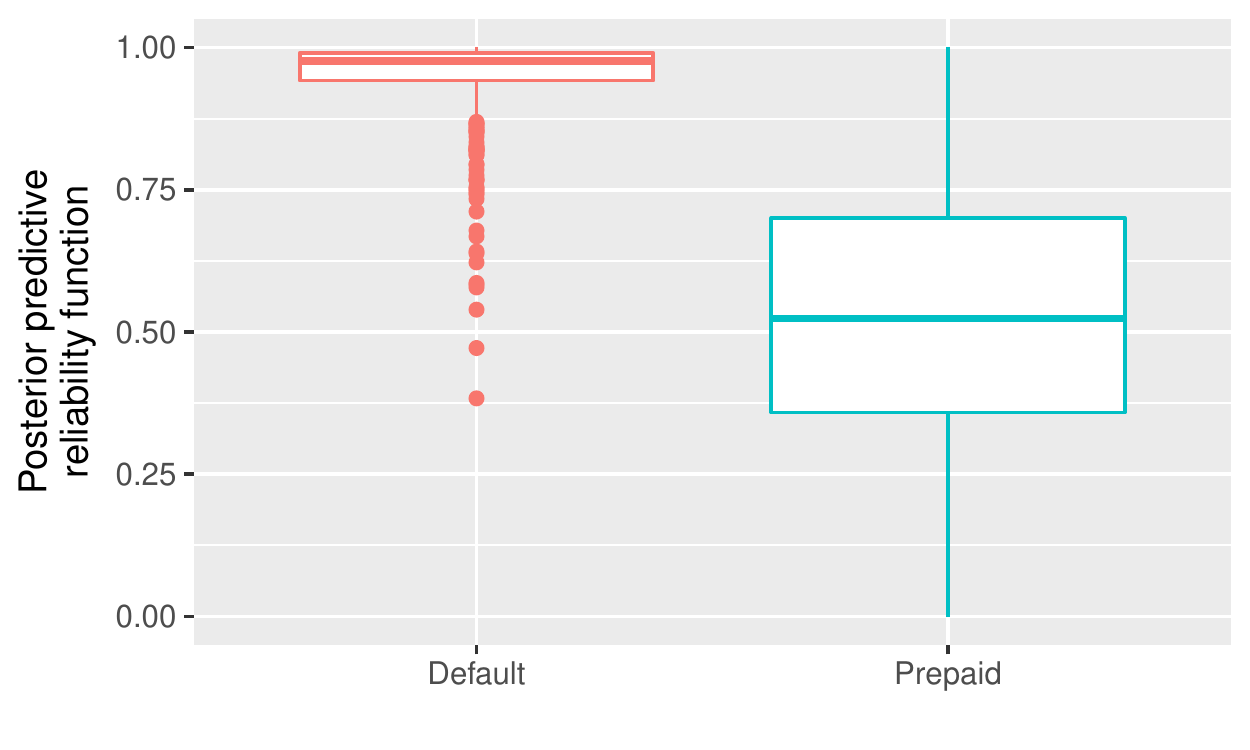}
\caption{Box plots of posterior predictive reliability function computed for mortgages at time to default and prepay.}
\label{fig:box_surv}
\end{figure}

In conclusion, one can say that the model fit has clearly identified the significant predictors affecting the mortgage status and also produced excellent prediction for prepaid mortgages. Because of the extremely disproportionate data in Freddie Mac, prediction for default mortgages was not as impressive as the prepaid ones.

%{\color{red} Do I need to write anything more? I can of course provide more plots or tables if you want. }

\section{Conclusion and future work}
In this paper we have introduced a model for the time to mortgage prepayment or default as a function of mortgage covariates.  The proposed competing risks model allows one to take account of the fact that an observation of a mortgage prepay is also a censored observation of a default, and vice versa; hence observation of one does contain information about the other that should be used in inference. Model inference can be done even for quite large data sets, as has been illustrated here for a set of single family loan data from Freddie Mac, where the relative effects of the different covariates on eventual prepayment or default have been quantified. Some difficulties with the inference were encountered, particularly for the defaults that were only a small percentage of the data.  In particular, the identifiability issues with this model can cause some convergence issues with the MCMC implementation of the inference.

Various extensions to the model are possible.  This model assumes that each covariate has the same effect on the prepay and default rate of every mortgage. Heterogeneity in these effects can be introduced through a Bayesian hierarchical model.
%where mortgage $i$ has its own parameters $\theta_{D,i}$, $\theta_{P,i}$ and $\psi_i$, and each parameter set is assumed to be a random sample from a population with some parameterized distribution e.g.\ $p(\psi_i \, | \, \Psi)$.  Inference can then be on the sets of individual-mortgage parameters and on the population level parameters such as $\Psi$. 
However our residual analysis did not detect any obvious clustering of residuals that would be indicative of such heterogeneity.

\section{References}
\bibliographystyle{agsm}
\bibliography{mortgage_bib}

\clearpage

\section{Appendix}
\appendix

\subsection*{Deriving the failure rate of the lognormal distribution}
Let $T$ be a lognormally distributed random variable with parameters $\mu$ and $\sigma^2$ and density function
\[ f(t \, | \, \mu, \sigma^2) = \frac{1}{\sqrt{2\pi\sigma^2}\, t} \exp\left(-\frac{1}{2\sigma^2} (\log(t)-\mu)^2\right). \] 
The failure rate is defined as
\[ r(t) = \frac{f(t \, | \, \mu, \sigma^2)}{P(T>t \, | \, \mu, \sigma^2)} = \frac{f(t \, | \, \mu, \sigma^2)}{\int_t^{\infty} f(s \, | \, \mu, \sigma_D) \: ds}. \]

The lognormal failure rate can be calculated in terms of the normal cdf  because $T$ has the property that $\log(T)$ is normally distributed. Therefore
\[ \int_t^{\infty} f(s \, | \, \mu, \sigma^2) \: ds = 1 - P(T < t) = 1 - P( \log(T) < \log(t)) = 1 - \Phi((\log (t) - \mu)/\sigma), \]
where $\Phi$ is the standard normal cdf.  Hence
\begin{equation} 
r(t) = \frac{(2\pi \sigma^2)^{-1/2} t^{-1} \exp\left(-0.5(\log (t) - \mu)^2 / \sigma^2 \right)}{1 - \Phi((\log (t) - \mu)/\sigma)}.
\label{eq:failurerate}
\end{equation}

\subsection*{Computing the integral of the failure rate function}
The integral of the failure rate function appears in the likelihood function.  It is assumed that the covariates $\bm{X}(t)$ vary piecewise constantly on intervals with mid-points $\tau_1 < \tau_2 < \cdots < \tau_m$. So $\bm{X}(t) = \bm{X}(\tau_j)$ for $s_{j-1} < t \leq s_j$, with interval end-points $s_0 =0$ and $s_j = 0.5(\tau_j + \tau_{j+1}), \: j=1,\ldots,m$, with $\tau_{m+1} = \infty$.

% Therefore $\{ X(w) \, | \, 0 \leq w \leq t\})$ is specified completely by $\bm{X}(t) = (X(\tau_1),X(\tau_2),\ldots,X(\tau_{m^{\prime}}))$, where   
Let $m^{\prime} = \max \{ j \, | \, \tau_j < t_D \}$. The integral of the failure rate, needed in the specification of the distribution of $T_D$, is then:
\begin{equation}
\int_0^{t_D} \lambda_D(w  \, | \, \bm{X}(w)) \: dw = \sum_{j=1}^{m^{\prime}} \exp(\theta_D^{\prime} \bm{X}(\tau_{j})) \int_{s_{j-1}}^{s_j} r_D(w) \: dw  \: + \:  \exp(\theta_D^{\prime} \bm{X}(\tau_{j})) \int_{s_{m^{\prime}}}^{t_D} r_D(w) \: dw
\label{eq:lambdaD_integral1}
\end{equation}
% and 
% \begin{equation}
% \int_0^{t_P} \lambda_P(w  \, | \, \bm{X}(t)) \: dw = \sum_{j=1}^{m^{\prime}-1} \exp(\theta_P^{\prime} X(\tau_{j})) \int_{\tau_{j}}^{\tau_{j+1}} r_P(w) \: dw  \: + \:  \exp(\theta_P^{\prime} X(\tau_{j})) \int_{\tau_{m^{\prime}}}^{t_P} r_P(w) \: dw.
% \label{eq:lambdaP_integral1}
% \end{equation}

The integral of the lognormal failure rate can be calculated in a closed form expression, using the fact that $\log(T)$ is Gaussian, and that
\[   - \log( P(T > t) ) =  \int_0^t r(s) \: ds \]
holds for any failure rate, so that:
\begin{eqnarray} 
\int_{t_a}^{t_b} r(s) \: ds & = & \int_{0}^{t_b} r(s) \: ds  -  \int_{0}^{t_a} r(s) \: ds \nonumber \\
& = & - \log( P(T > t_b) ) + \log( P(T > t_a) ) \nonumber \\
& = & - \log( 1 - \Phi[(\log (t_b) - \mu)/\sigma]) + \log(1 -  \Phi[(\log (t_a) - \mu)/\sigma]). 
\label{eq:intrate}
\end{eqnarray}

Substituting Equation \ref{eq:intrate} into Equation \ref{eq:lambdaD_integral1} gives:
\begin{multline}
\int_0^{t_D} \lambda_D(w \, | \, \bm{X}(w)) \: dw \\ 
= \: \sum_{j=1}^{m^{\prime}} \exp(\theta_D^{\prime} \bm{X}(\tau_{j})) \Biggl[ - \log( 1 - \Phi[(\log (s_{j}) - \mu_D)/\sigma_D]) + \log( 1 - \Phi[(\log (s_{j-1}) - \mu_D)/\sigma_D]) \Biggr]  \\ 
+ \:  \exp(\theta_D^{\prime} \bm{X}(\tau_{m^{\prime}})) \Biggl[ - \log(1 - \Phi[(\log (t_D) - \mu_D)/\sigma_D]) + \log( 1 - \Phi[(\log (s_{m^{\prime}}) - \mu_D)/\sigma_D]) \Biggr]
\label{eq:lambdaD_integral2}
\end{multline}
% and 
% \begin{multline}
% \int_0^{t_P} \lambda_P(w \, | \, X_i) \: dw 
% \\ = \:\sum_{j=1}^{m^{\prime}-1} \exp(\theta_P^{\prime} X_i(\tau_{j})) \Biggl[ - \log( 1 - \Phi[(\log (\tau_{j+1}) - \mu_P)/\sigma_P]) + \log( 1 - \Phi[(\log (\tau_j) - \mu_P)/\sigma_P]) \Biggr]  \\ 
% + \:  \exp(\theta_P^{\prime} X_i(\tau_{m^{\prime}})) \Biggl[ - \log(1 - \Phi[(\log (t_D) - \mu_P)/\sigma_P]) + \log( 1 - \Phi[(\log (t_{m^{\prime}}) - \mu_P)/\sigma_P]) \Biggr]\label{eq:lambdaP_integral2}
% \end{multline}
The integral for $T_P$, $\int_0^{t_P} \lambda_P(w \, | \, \bm{X}(w)) \: dw$ is also given by Equation \ref{eq:lambdaD_integral2} with $t_D$, $\theta_D, \mu_D$ and $\sigma_D$ replaced by $t_P$, $\theta_P, \mu_P$ and $\sigma_P$ respectively.

\subsection*{Details of the MCMC Algorithm for the Homogeneous Model}
Sampling of the posterior distribution of Equation \ref{eq:posterior}, with likelihood given by Equation \ref{eq:likelihood}, is done by a Metropolis within Gibbs algorithm. Each block of parameters are sampled from their full conditional distribution, with those samples obtained through a Metropolis proposal, as follows:

\paragraph{Sample $\bm{\theta_D}$}
From a current value $\bm{\theta_D}$, a random walk proposal $\theta_D^*$ is made from a Gaussian with mean $\theta_D$ and variance $s_{\theta, D}^2 I_{m \times m}$, where $I_{m \times m}$ is the identity matrix of dimension $m$ and $s_D^2$ is tuned to provide a reasonable acceptance rate. The proposal is accepted with probability
\begin{multline*}
\min \left\{1, \frac{p(\bm{t}_D, \bm{t}_P, \bm{t}_C \, | \, \theta_D^*, \theta_P, \psi, {\cal X}) \: p(\theta_D^*)}{p(\bm{t}_D, \bm{t}_P, \bm{t}_C \, | \, \theta_D, \theta_P, \psi, {\cal X}) \: p(\theta_D)} \right\} \; = \;
\min \left\{1, \frac{p(\theta_D^*) \: \prod_{i=1}^{n_D} \lambda_D^*(t_i^D \,|\,\bm{X}_i(t_i^D))}{p(\theta_D) \: \prod_{i=1}^{n_D} \lambda_D(t_i^D \,|\,\bm{X}_i(t_i^D))} \right. \\ 
\left. \times \: \frac{\exp\left( -\sum_{i=1}^{n_D} \int_0^{t_i^D} \lambda_D^*(w \, | \, \bm{X}_i(w)) \: dw -  \sum_{i=n_D+1}^{n_D+n_P} \int_0^{t_i^P} \lambda_D^*(w \, | \, \bm{X}_i(w)) \: dw - \sum_{i=n_D+n_P+1}^{N} \int_0^{t_i^C} \lambda_D^*(w \, | \, \bm{X}_i(w)) \: dw \right)}{\exp\left( -\sum_{i=1}^{n_D} \int_0^{t_i^D} \lambda_D(w \, | \, \bm{X}_i(w)) \: dw -  \sum_{i=n_D+1}^{n_D+n_P} \int_0^{t_i^P} \lambda_D(w \, | \, \bm{X}_i(w)) \: dw - \sum_{i=n_D+n_P+1}^{N} \int_0^{t_i^C} \lambda_D(w \, | \, \bm{X}_i(w)) \: dw \right)} \right\},
\end{multline*}
where: $\lambda_D^*(t \, | \, \bm{X}(t))$ is given by Equation \ref{eq:failurerateD} with $\theta_D = \theta_D^*$, $\bm{X}(t)$ is given by Equation \ref{eq:piecewise} and $\int_0^t \lambda_D(w \, | \, X(w)) \:dw$ is given by Equation \ref{eq:lambdaD_integral2}.

\paragraph{Sample $\bm{\theta_P}$}
This is identical to sampling from $\theta_D$, with $\lambda_D(t \, | \bm{X}(t))$ replaced by $\lambda_P(t \, | \bm{X}(t))$ throughout.

\paragraph{Sample $\bm{\mu}_D$}
From a current value $\mu_D$, a random walk proposal $\mu_D^*$ is made from a Gaussian with mean $\mu_D$ and variance $s_{\mu,D}^2$, where $s_{\mu,D}^2$ is tuned to provide a reasonable acceptance rate. The proposal is accepted with probability
\begin{multline}
\min \left\{1, \frac{p(\bm{t}_D, \bm{t}_P, \bm{t}_C \, | \, \theta_D, \theta_P, \psi^*, {\cal X}) \: p(\mu_D^*)}{p(\bm{t}_D, \bm{t}_P, \bm{t}_C \, | \, \theta_D, \theta_P, \psi, {\cal X}) \: p(\mu_D)} \right\} \; = \;
\min \left\{1, \frac{p(\mu_D^*) \: \prod_{i=1}^{n_D} \lambda_D^*(t_i^D \,|\,\bm{X}_i(t_i^D))}{p(\mu_D) \: \prod_{i=1}^{n_D} \lambda_D(t_i^D \,|\,\bm{X}_i(t_i^D))} \right. \\ 
\left. \times \: \frac{\exp\left( -\sum_{i=1}^{n_D} \int_0^{t_i^D} \lambda_D^*(w \, | \, \bm{X}_i(w)) \: dw -  \sum_{i=n_D+1}^{n_D+n_P} \int_0^{t_i^P} \lambda_D^*(w \, | \, \bm{X}_i(w)) \: dw - \sum_{i=n_D+n_P+1}^{N} \int_0^{t_i^C} \lambda_D^*(w \, | \, \bm{X}_i(w)) \: dw \right)}{\exp\left( -\sum_{i=1}^{n_D} \int_0^{t_i^D} \lambda_D(w \, | \, \bm{X}_i(w)) \: dw -  \sum_{i=n_D+1}^{n_D+n_P} \int_0^{t_i^P} \lambda_D(w \, | \, \bm{X}_i(w)) \: dw - \sum_{i=n_D+n_P+1}^{N} \int_0^{t_i^C} \lambda_D(w \, | \, \bm{X}_i(w)) \: dw \right)} \right\},
\label{eq:muD_sample}
\end{multline}
where: $\psi^* = (\mu_D^*,\sigma_D^2,\mu_P,\sigma_P^2)$, $\lambda_D^*(t \, | \, \bm{X}(t))$ is given by Equation \ref{eq:failurerateD} with $\mu_D = \mu_D^*$, $\bm{X}(t)$ is given by Equation \ref{eq:piecewise} and $\int_0^t \lambda_D(w \, | \, X(w)) \:dw$ is given by Equation \ref{eq:lambdaD_integral2}.

\paragraph{Sample $\bm{\mu_P}$}
This is identical to sampling from $\mu_D$, with $\lambda_D(t \, | \bm{X}(t))$ replaced by $\lambda_P(t \, | \bm{X}(t))$ throughout and $\psi^* = (\mu_D,\sigma_D^2,\mu_P^*,\sigma_P^2)$.

\paragraph{Sample $\bm{\sigma_D^2}$}
From a current value $\sigma_D^2$, a proposal $\sigma_D^{2,*}$ is generated from a uniform distribution on the interval $(a \sigma_D^2, \sigma_D^2/a)$, where $a \in (0,1)$ is tuned to provide a reasonable acceptance rate. The proposal is accepted with probability
\begin{multline}
\min \left\{1, \frac{p(\bm{t}_D, \bm{t}_P, \bm{t}_C \, | \, \theta_D, \theta_P, \psi^*, {\cal X}) \: p(\sigma_D^{*,2}) \: p(\sigma_D^2 \, | \, \sigma_D^{2,*})}{p(\bm{t}_D, \bm{t}_P, \bm{t}_C \, | \, \theta_D, \theta_P, \psi, {\cal X}) \: p(\sigma_D^2) \: p(\sigma_D^{2,*} \, | \, \sigma_D^2)} \right\} \; = \;
\min \left\{1, \frac{\sigma_D^2 \: p(\sigma_D^{2,*}) \: \prod_{i=1}^{n_D} \lambda_D^*(t_i^D \,|\,\bm{X}_i(t_i^D))}{\sigma_D^{2,*} p(\sigma_D^2) \: \prod_{i=1}^{n_D} \lambda_D(t_i^D \,|\,\bm{X}_i(t_i^D))} \right. \\ 
\left. \times \: \frac{\exp\left( -\sum_{i=1}^{n_D} \int_0^{t_i^D} \lambda_D^*(w \, | \, \bm{X}_i(w)) \: dw -  \sum_{i=n_D+1}^{n_D+n_P} \int_0^{t_i^P} \lambda_D^*(w \, | \, \bm{X}_i(w)) \: dw - \sum_{i=n_D+n_P+1}^{N} \int_0^{t_i^C} \lambda_D^*(w \, | \, \bm{X}_i(w)) \: dw \right)}{\exp\left( -\sum_{i=1}^{n_D} \int_0^{t_i^D} \lambda_D(w \, | \, \bm{X}_i(w)) \: dw -  \sum_{i=n_D+1}^{n_D+n_P} \int_0^{t_i^P} \lambda_D(w \, | \, \bm{X}_i(w)) \: dw - \sum_{i=n_D+n_P+1}^{N} \int_0^{t_i^C} \lambda_D(w \, | \, \bm{X}_i(w)) \: dw \right)} \right\},
\label{eq:sigmaD_sample}
\end{multline}
where: $\psi^* = (\mu_D,\sigma_D^{2,*},\mu_P,\sigma_P^2)$, $\lambda_D^*(t \, | \, \bm{X}(t))$ is given by Equation \ref{eq:failurerateD} with $\sigma_D^2 = \sigma_D^{2,*}$, $\bm{X}(t)$ is given by Equation \ref{eq:piecewise} and $\int_0^t \lambda_D(w \, | \, X(w)) \:dw$ is given by Equation \ref{eq:lambdaD_integral2}.

\paragraph{Sample $\bm{\sigma_P^2}$}
This is identical to sampling from $\sigma_D^2$, with $\lambda_D(t \, | \bm{X}(t))$ replaced by $\lambda_P(t \, | \bm{X}(t))$ throughout and $\psi^* = (\mu_D,\sigma_D^2,\mu_P,\sigma_P^{2,*})$.

\subsection*{MCMC output plots}
\begin{figure}[!h]
\centering
\captionsetup{justification=raggedright}
\includegraphics{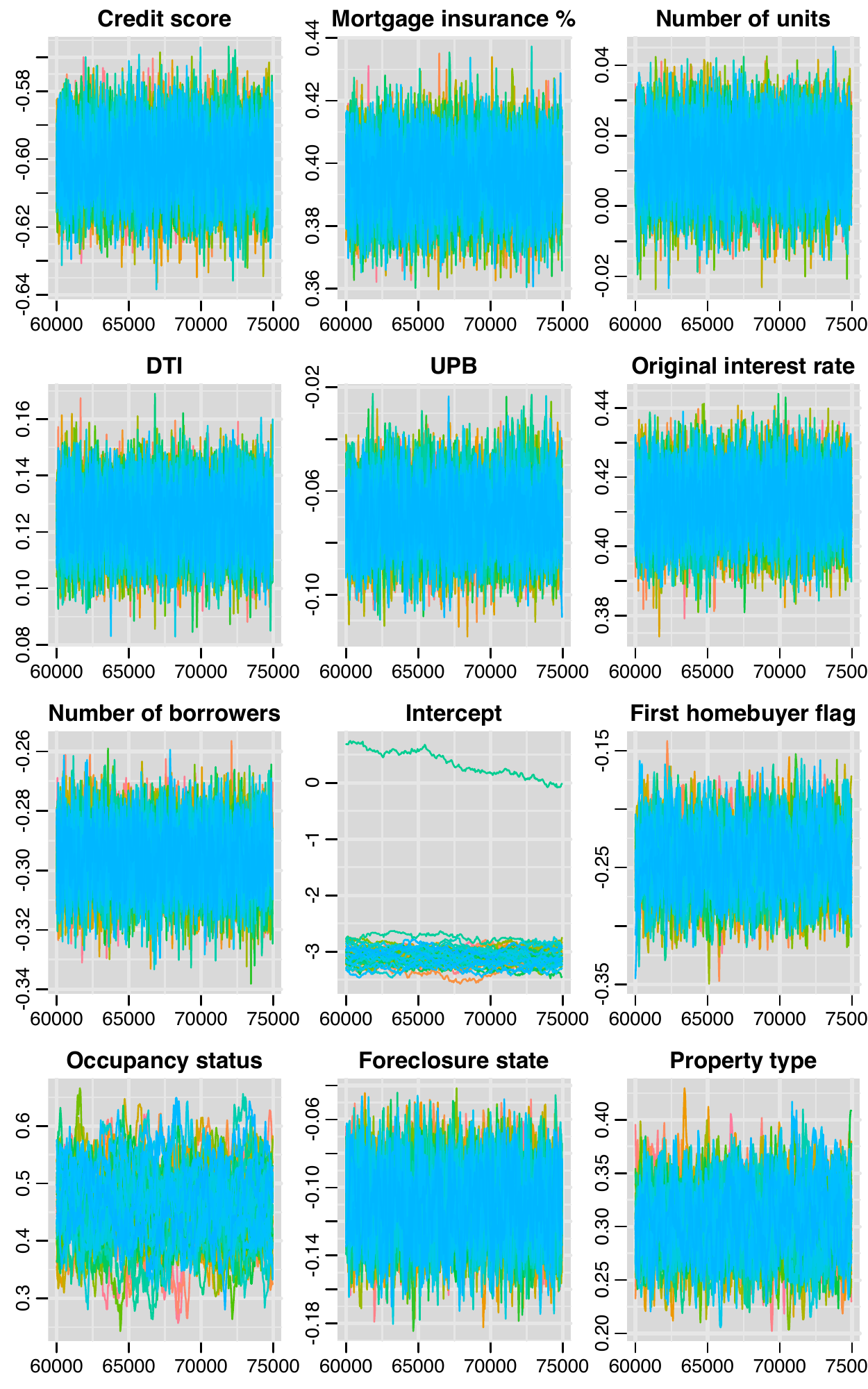}
\caption{Trace plot of all parameters associated with covariates for default category. A single chain for intercept parameter is found to converge much more slowly than the others.}
\label{fig:trace_covar_default}
\end{figure}

\begin{figure}[!h]
\centering
\captionsetup{justification=raggedright}
\includegraphics{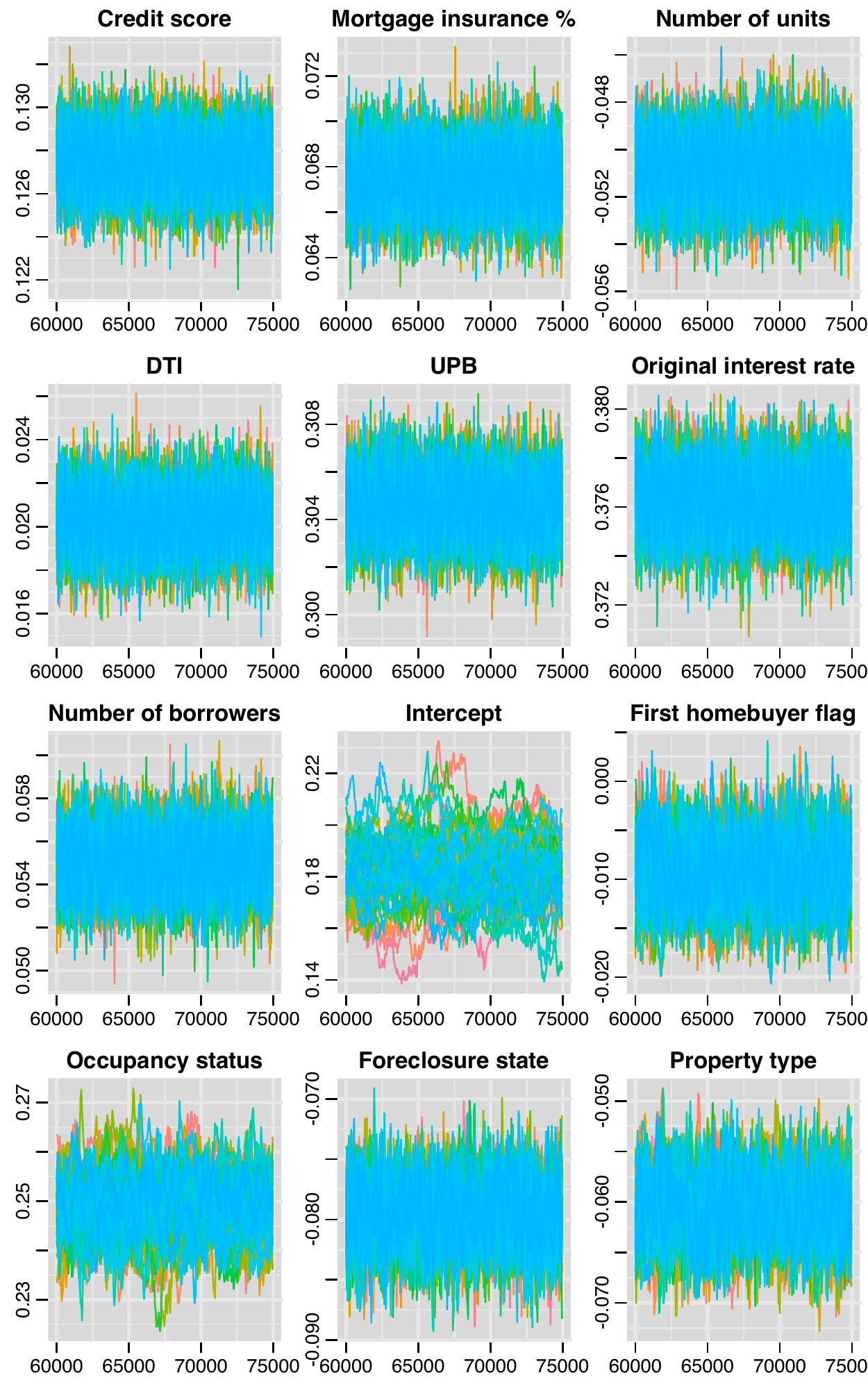}
\caption{Trace plot of all parameters associated with covariates for prepaid category.}
\label{fig:trace_covar_prepaid}
\end{figure}

\begin{figure}[!h]
\centering
\captionsetup{justification=raggedright}
\includegraphics{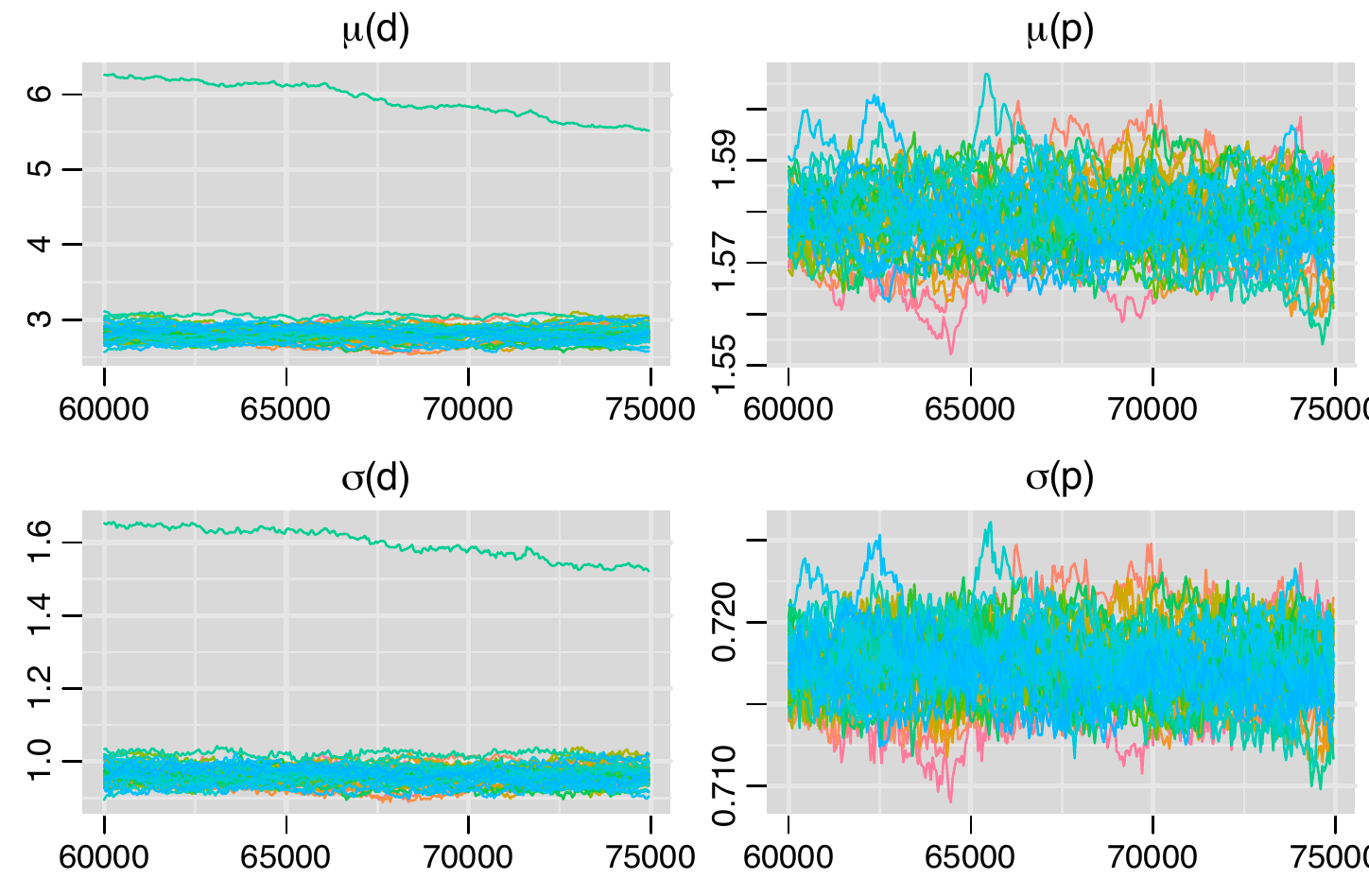}
\caption{Trace plot of all distributional parameters. For parameters in default category, the single slow converging chain is visible here as well.}
\label{fig:trace_distri_par_all}
\end{figure}

The trace plots for all the variables for category default are provided in figure \ref{fig:trace_covar_default} which seems to indicate seem to converge fairly well. The problem with a single chain is noticeable in the intercept ($\beta_0$) and distributional mean ($\mu$) and s.d.\ ($\sigma$) traces, which can possibly be attributed to the identifiability issue discussed earlier.
%\begin{centering}
%\end{centering}
Trace plots of parameters associated with category prepaid are provided in figure \ref{fig:trace_covar_prepaid}. The traces converge well and see to have identified the posteriors satisfactorily.
Finally trace plots of distributional parameters $\mu_d, \mu_p,\sigma_d,\sigma_p$ are provided in \ref{fig:trace_distri_par_all}. A single slow converging chain is again found in the default category parameters. A larger proportion of default mortgages data  and/or a longer run of the chain would have prevented this problem.

%Caterpillar plot of all the parameters are given in \ref{fig:cater_allpar}. The thick line represent 95\% and the thin line 68\% HPD intervals.
%\begin{centering}
%\begin{figure}[!h]
%\captionsetup{justification=centering}
%\includegraphics{caterpillar_plot}
%\caption{Caterpillar plot of all parameters.}
%\label{fig:cater_allpar}
%\end{figure}
%\end{centering}
%Finally we provide trace plot of all the parameters in figure \ref{fig:trace_allpar}. The bad mixing of prepay parameters are evident.
%\begin{centering}
%\begin{figure}[!h]
%\includegraphics{all_trace.pdf}
%\caption{Trace plot of all parameters.}
%\label{fig:trace_allpar}
%\end{figure}
%\end{centering}
%%%%%%%%%%%%%%%%%%%%%%%%%%%%%%%%%%%%%%%%%%%%%%%%%%%%%%%%%%%%%%%%%%%
%%%%%%%%%%%%%%%%%%%%%%%%%%%%%%%%%%%%%%%%%%%%%%%%%%%%%%%%%%%%%%%%%%%

\end{document}